\documentclass[sigconf]{acmart}

\usepackage{balance} 
\usepackage{multirow}
\AtBeginDocument{%
  }

\copyrightyear{2023}
\acmYear{2023}
\setcopyright{rightsretained}
\acmConference[HRI '23]{Proceedings of the 2023 ACM/IEEE International Conference on Human-Robot Interaction}{March 13--16, 2023}{Stockholm, Sweden}
\acmBooktitle{Proceedings of the 2023 ACM/IEEE International Conference on Human-Robot Interaction (HRI '23), March 13--16, 2023, Stockholm, Sweden}\acmDOI{10.1145/3568162.3576987}
\acmISBN{978-1-4503-9964-7/23/03}

\begin{document}

\title{Evaluating and Personalizing User-Perceived Quality of Text-to-Speech Voices for Delivering Mindfulness Meditation with Different Physical Embodiments}

\author{Zhonghao Shi}
\authornote{Both authors contributed equally to this research.}

\author{Han Chen}
\authornotemark[1]
\affiliation{%
  \institution{University of Southern California}
  \city{Los Angeles}
  \state{CA}
  \country{USA}
}

\author{Anna-Maria Velentza}
\affiliation{%
  \institution{University of Southern California}
  \city{Los Angeles}
  \state{CA}
  \country{USA}
}

\author{Siqi Liu}
\affiliation{%
  \institution{University of Southern California}
  \city{Los Angeles}
  \state{CA}
  \country{USA}
}

\author{Nathaniel Dennler}
\affiliation{%
  \institution{University of Southern California}
  \city{Los Angeles}
  \state{CA}
  \country{USA}
}

\author{Allison O'Connell}
\affiliation{%
  \institution{University of Southern California}
  \city{Los Angeles}
  \state{CA}
  \country{USA}
}

\author{Maja Mataric}
\affiliation{%
  \institution{University of Southern California}
  \city{Los Angeles}
  \state{CA}
  \country{USA}
}






\renewcommand{\shortauthors}{Zhonghao Shi et al.}

\begin{abstract}

Mindfulness-based therapies have been shown to be effective in improving mental health, and technology-based methods have the potential to expand the accessibility of these therapies. To enable real-time personalized content generation for mindfulness practice in these methods, high-quality computer-synthesized text-to-speech (TTS) voices are needed to provide verbal guidance and respond to user performance and preferences. However, the user-perceived quality of state-of-the-art TTS voices has not yet been evaluated for administering mindfulness meditation, which requires emotional expressiveness. In addition, work has not yet been done to study the effect of physical embodiment and personalization on the user-perceived quality of TTS voices for mindfulness. To that end, we designed a two-phase human subject study. In Phase 1, an online Mechanical Turk between-subject study (N=471) evaluated 3 (feminine, masculine, child-like) state-of-the-art TTS voices with 2 (feminine, masculine) human therapists' voices in 3 different physical embodiment settings (no agent, conversational agent, socially assistive robot) with remote participants. Building on findings from Phase 1, in Phase 2, an in-person within-subject study (N=94), we used a novel framework we developed for personalizing TTS voices based on user preferences, and evaluated user-perceived quality compared to best-rated non-personalized voices from Phase 1. We found that the best-rated human voice was perceived better than all TTS voices; the emotional expressiveness and naturalness of TTS voices were poorly rated, while users were satisfied with the clarity of TTS voices. Surprisingly, by allowing users to fine-tune TTS voice features, the user-personalized TTS voices could perform almost as well as human voices, suggesting user personalization could be a simple and very effective tool to improve user-perceived quality of TTS voice.
\end{abstract}


\maketitle

\section{INTRODUCTION}

\begin{figure}[t!]
  \centering
  \includegraphics[scale=0.32]{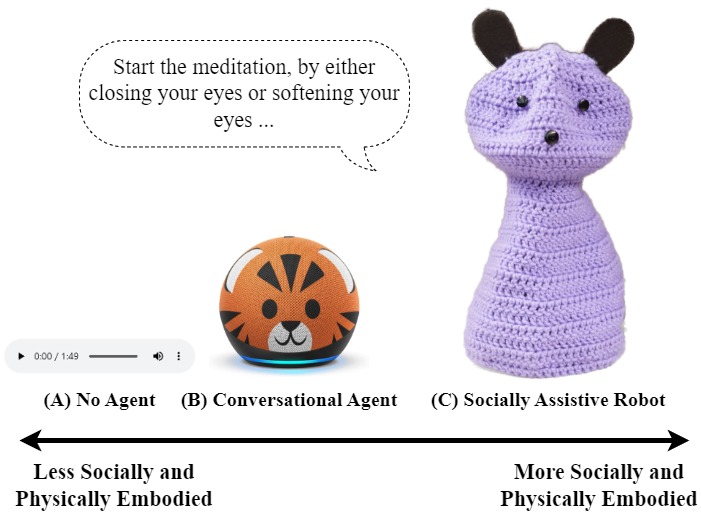}
  \caption{The three embodiments studied}
\label{fig:embodi}
\end{figure}

Mindfulness-based cognitive therapies have been shown to be effective in reducing stress, anxiety, and depression~\cite{toneatto2007does}. The practice of mindfulness requires self-discipline, making adherence challenging. In-person guided mindfulness practice reinforces adherence and efficacy, but accessibility and affordability of in-person sessions with human therapists are not scalable~\cite{barnard2020understanding}. To expand access, past work has studied technology-based methods for administering mindfulness practice, such as mobile applications (apps)~\cite{flett2019mobile}, conversational agents~\cite{gardiner2017engaging}, and socially assistive robots~\cite{sliwinski2017review}. Such methods largely used pre-recorded guided meditation scripts and therefore offered limited or no real-time support or feedback to the user~\cite{dauden2018evaluating}. In contrast, TTS voices can be amended and synthesized in real-time, allowing for personalization of content based on each user's needs. Therefore, to enable better technology-based mindfulness therapies, high-quality synthesized text-to-speech (TTS) voices are needed to deliver mindfulness guidance with the level of quality and efficacy approaching human therapists in different physically embodied settings.

In the speech synthesis community, TTS voices have been commonly evaluated for their clarity~\cite{cambre2020choice, bennett2005large, cambre2019one, king2014measuring, wagner2019speech} and intelligibility (i.e., how easy it is to accurately hear the content of the speech)~\cite{wagner2019speech, govender2018using, king2014measuring} in the context of reading. With the recent progress in TTS, studies have shown that TTS voices are highly effective at communicating textual information in these two aspects~\cite{cambre2020choice, wagner2019speech}. However, there has not yet been much evaluation of the user-perceived quality of TTS voices on socio-emotional tasks--like mindfulness meditation--that require emotional expressiveness. Furthermore, past work has already suggested the impact of voice on patient satisfaction in the medical and mental health care settings~\cite{haskard2008provider, liu2020physician}. For these reasons, {\it our work aims to address the research question of how well state-of-the-art TTS voices perform in mindfulness-based meditation interventions, which require the voice to not only clearly communicate textual information but also to accurately convey calming emotional support.} Our work evaluated the performance of three state-of-the-art TTS voices compared with human therapists' voices to inform future TTS voice selection for technology-based mindfulness practice.

Past work has shown great success in delivering mindfulness with a diverse set of devices and agents. This opens up the research question of the effect of the agent's physical embodiment on the user-perceived quality of TTS for mindfulness. Prior work has suggested that a mismatch between the visual embodiment and voice of a robot or agent can cause feelings of uneasiness~\cite{mitchell2011mismatch}. Other work has shown that there is no one voice that fits all physical embodiments~\cite{cambre2019one}, and people reliably associate a robot's physical embodiment with their expectation of the robot's voice. Past research has also indicated that some voices may align better with certain physical embodiments than others~\cite{mcginn2019can}. These findings motivated us to hypothesize and investigate the effects of different physical embodiments on user-perceived TTS voice quality for mindfulness.

Previous studies have indicated that the level of alignment between the user's background and the characteristics of the agent's voice may affect the experience of human-machine or human-robot interaction~\cite{cambre2019one}. Works in HRI have further demonstrated that adapting the features of a robot's voice to a user's voice improves rapport and trustworthiness~\cite{lubold2016effects}. In addition, past literature has shown that offering users opportunities to choose and have control may help improve user perceptions of chosen experiences~\cite{leotti2011inherent, leotti2010born}. These insights led us to hypothesize that personalization of TTS voice features may also improve user-perceived voice quality for mindfulness meditation. Our work studied the effect of personalization on TTS voice in the context of mindfulness.


 

Through a two-phase human subject study, we evaluated \textcolor{red}{}{user-perceived quality of state-of-the-art TTS voices} for mindfulness meditation practice in three embodiment conditions (Phase 1), and explored the effect of personalization on TTS voices (Phase 2). As detailed in Table A of the supplemental material, in Phase 1, a Mechanical Turk (MTurk) study evaluated and compared 3 state-of-the-art TTS voices (feminine, masculine, child-like) with 2 prerecorded human therapist voices (feminine, masculine). As shown in Figure~\ref{fig:embodi}, we evaluated 3 physical embodiment conditions (no agent, conversational agent, socially assistive robot) to study the effect of physical embodiment on the user's perception of voice. In Phase 2, we conducted an in-person study that compared three conditions: 1) the best-performing TTS voice (from Phase 1); 2) the best-performing human therapist's voice (also from Phase 1); 3) a personalized voice fine-tuned by users.

Our findings show that the best-rated human therapist's voice was perceived as significantly better than all the TTS voices for mindfulness meditation. Users were generally satisfied with TTS voice clarity, but gave significantly more negative ratings to the emotional expressiveness and naturalness of TTS voices. Furthermore, we found that physical embodiment had an effect on one TTS voice, but not on the other TTS and human voices. Unexpectedly, we found the alignment between the robot embodiment and voice may help to remind and amplify the user’s dislike of the artificial sound of a TTS voice. Finally, we showed that user-personalized TTS voices were rated significantly better than the non-personalized TTS voices, and almost as well as human voices. This may suggest that personalization can be a simple but effective participatory design strategy to improve the user-perceived quality of TTS voices.

Following gender guidelines \cite{cordero2022and, scheuerman2020hci} and recognized effects of embodiment on perceived gender in robots \cite{perugia2022shape}, in the context of this study we refer to voices by their characteristics (masculine, feminine). This paper is organized as follows. In Section 2, we review past work done in the two research areas most related to this work, and discuss research gaps our work addresses. Section 3 includes a detailed methodology of our two-phase user study to support replication of this work. Section 4 presents all the study findings toward the understanding of the user-perceived quality of TTS voices for mediating mindfulness meditation, and the effect of physical embodiment and personalization on TTS voices. Section 5 discusses the contributions and significance of our findings for HRI, and Section 6 concludes the paper.

\section{Related Work}

\subsection{Text-to-Speech (TTS) voice evaluation}

Recent advances in machine learning show great progress in synthesizing TTS voices with high clarity~\cite{cambre2020choice, bennett2005large, cambre2019one, king2014measuring, wagner2019speech} and intelligibility (i.e., how easy it is to accurately hear the content of the speech)~\cite{wagner2019speech, govender2018using, king2014measuring}. State-of-the-art TTS voices currently perform close to human levels in functional aspects of text communication. Most recent large-scale studies have evaluated the state-of-the-art TTS voices for reading long-form content~\cite{cambre2020choice}. The results showed that, overall, the best-performed TTS voices performed almost as well as human voices, but no single voice outperformed the others across all evaluation dimensions such as clarity and naturalness. Some studies have suggested that the intelligibility level of the state-of-the-art TTS voices has already achieved near-human performance~\cite{wagner2019speech, govender2018using, king2014measuring}. 

However, designing TTS voices that are appropriate for individual users' preferences in specific application domains such as mindfulness meditation remains an open challenge~\cite{trouvain2020speech}, in part because studies have shown that current TTS voices are not as emotionally expressive as human voices~\cite{cohn2020differences, turk2010evaluation}, making it difficult to convey social aspects of human interaction, such as feelings and empathy, in synthesized speech. A recent study showed that Amazon Alexa voices were still very limited in their ability to convey accurate emotions compared to human voices~\cite{cohn2020differences}. Another study showed that there is a tradeoff between emotional expressiveness and naturalness, and the current emotionally expressive TTS voices cannot yet strike the balance between the two~\cite{turk2010evaluation}. 

Since no prior work to our knowledge has evaluated TTS voices for mindfulness meditation, we believe there is a gap in understanding how well the current state-of-the-art TTS voices would perform on tasks like mediating mindfulness practice that requires appropriate emotional expressiveness. 

\subsection{Technology-Mediated Mindfulness Practice}
Mindfulness exercises have been shown to be beneficial for improving cognitive performance, well-being, and mental state~\cite{gu2015mindfulness}. Many technology-based methods have been studied and developed to supplement and augment the efficacy of mindfulness-based therapies so that they can be accessible to users with diverse socio-economic backgrounds. They can be categorized into three groups: 1) No agent: past studies have shown that mindfulness meditation mobile apps such as Calm and Headspace can help to alleviate stress and support positive mental health if used regularly~\cite{flett2019mobile}; 2) Conversational agents: embodied smart speakers have been shown to be beneficial for well-being and mental health support~\cite{gardiner2017engaging}; 3) Socially assistive robots: a robot therapist's physical embodiment has been shown to create an enjoyable interaction that enhances user motivation and performance~\cite{tapus2009role}.

Despite the accessibility, convenience, and benefits of mobile apps, studies have shown that adhering to daily mindfulness practice is challenging, and embodied agents like socially assistive robots can help improve adherence and engagement level~\cite{ostrowski2021small}. A study showed that participants completing a relaxation exercise led by a robotic agent demonstrated more positive emotions, participation, and engagement than participants completing the same exercise led by a tablet, indicating the importance of the physical embodiment of the agent~\cite{mann2015people}. In another study, participants engaged in mindfulness meditation practice with a robot or by listening to a lecture with the robot’s voice while their brain activity was recorded. EEG findings suggested that the group meditating with the robot achieved a mindful state and reduced cognitive processing better than the group in the audio condition~\cite{yoon2021effect}.

Since different physical embodiments have advantages in different contexts, we explored the user-perceived quality of TTS voices of different physical embodiments in the context of mindfulness meditation, and how we might optimize the users' listening experience of a voice in different embodiments.

\section{Methods}
\subsection{Research Hypotheses}
This work investigates the following three research hypotheses:

\begin{enumerate}
    \item[H1:] Users will rate text-to-speech (TTS) voices lower than the best human therapists' voices in the context of mindfulness meditation.
    \item[H2:] Users will rate the same TTS voice differently (i.e., significantly lower or higher) depending on the embodiment
    \item[H3:] Users will rate user-personalized TTS voices higher compared to the non-personalized TTS voices. 
\end{enumerate}

\begin{figure}[h]
  \centering
  \includegraphics[scale=0.28]{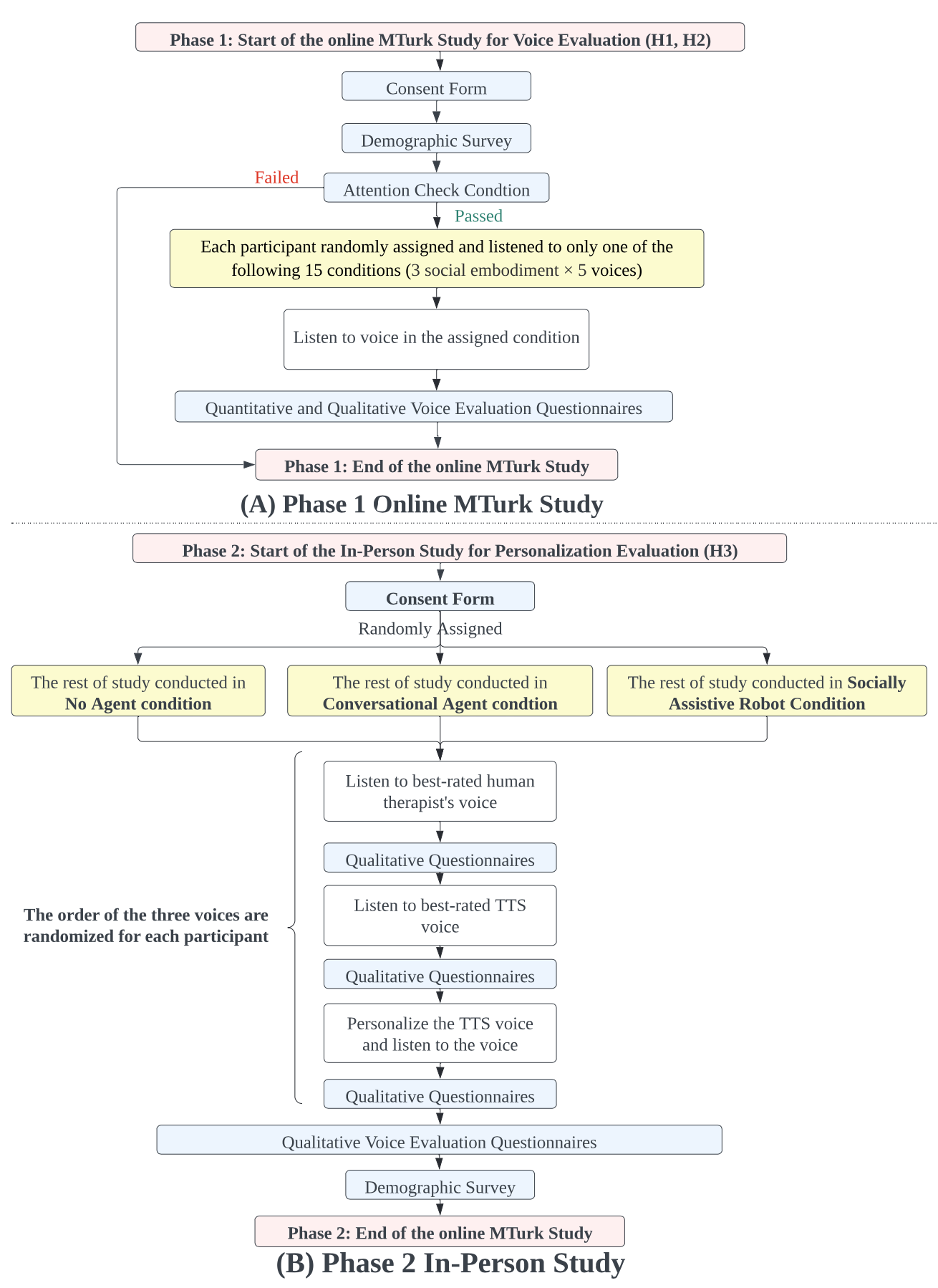}
  \caption{The two-phase study procedure: (A) phase 1 online MTurk study; (B) phase 2 in-person study}
\label{fig:flow}
\end{figure}

\subsection{Study Design}

To explore the above research hypotheses, we designed a two-phase study with two separate cohorts of participants, as shown in Figure~\ref{fig:flow}.

\subsubsection{Phase 1: Online MTurk Study for Voice Evaluation} 

Based on the evaluation framework introduced in~\cite{cambre2020choice}, we utilized a between-subject study design (N = 471) to investigate the first two research hypotheses (H1, H2). As shown in Figure \ref{fig:flow} (A), we developed an online web application that delivers a 2-minute mindfulness practice with one of 3 different embodiment conditions (no agent, conversational agent, socially assistive robot) and one of the 5 voice conditions (human feminine(HF), human masculine (HM), TTS feminine (SF), TTS masculine (SM), TTS child-like (SC)). The 3 physical embodiment conditions and 5 voice conditions formed 15 different study conditions; each participant was randomly assigned to one of the 15 conditions. Prior work indicated that reliable results can only be reached with at least 30 listeners in TTS evaluation~\cite{wester2015listener}. Following that guideline, we recruited at least 30 participants for each condition to ensure sufficient data for reliable results.

\subsubsection{Phase 2: In-Person Study for Personalization Evaluation}

We conducted a mixed-design study with counterbalancing to explore the third research hypothesis (H3) on personalization. As shown in Figure \ref{fig:flow} (B), participants were recruited and randomly assigned into 3 groups corresponding to each of the physical embodiment conditions. According to the literature~\cite{wester2015listener}, to reach a stable level of significance, we recruited at least 30 participants for each condition. The best-rated human therapist's voice and the best-rated TTS voice from each embodiment condition reported in Phase 1 were used as the non-personalized baselines in Phase 2. Instead of watching a video recording online, in Phase 2 participants were invited to the lab and asked to follow the same 2-min mindfulness practice in person. To eliminate potential ordering effects, the three types of voices (two non-personalized and one user-personalized) were presented in randomized order.

\subsection{Procedure}
Both phases of the study were approved by the Institutional Review Board at the University of Southern California (ID: UP-21-00984).

\subsubsection{Phase 1: Online MTurk Study for Voice Evaluation} 
As shown in Figure \ref{fig:flow} (A), in Phase 1, after the participants accepted the task, they were first asked to review the consent information sheet, and fill out the demographic survey. We also included a timer that served as an attention check. Participants who completed the survey in less than 2 minutes did not pass the attention check and were excluded from the data analysis. The qualified participants were then randomly assigned to one of the 15 participant conditions. Based on the condition they were assigned to, the participants were asked to participate in 2-min mindfulness practice, while paying attention to the quality of the voice guiding the practice. After the practice, participants were asked to complete a set of quantitative and qualitative questionnaires about their perceived ratings of the voice.

\subsubsection{Phase 2: In-Person Study for Personalization Evaluation}
As shown in Figure \ref{fig:flow} (B), participants were invited to the research lab and asked to review the consent information sheet. The participants were randomly assigned into one of 3 groups corresponding to the physical embodiment conditions. The participants were then asked to listen to the same 2-min mindfulness practice, and evaluate the following three voices in randomized order: 1) best-rated human therapist's voice in this embodiment condition from Phase 1; 2) best-rated TTS voice in this embodiment condition from Phase 2; 3) user's personalized TTS voice. The first two voice conditions were used as the non-personalized baselines. In the third voice condition, a personalization panel was introduced to allow participants to manually fine-tune voice features (accent, gender, speed, pitch and break time) based on their preferences. The participants were asked to use this process of personalization repeatedly until they obtained the desired voice. After the evaluation of three voices, the participants were also asked to complete a qualitative questionnaire and a demographic survey.

\subsection{Participants}
\subsubsection{Phase 1: Online MTurk Study for Voice Evaluation} 
As shown in Table A of the supplemental material, in Phase 1, we recruited participants (N=471) from Amazon Mechanical Turk (MTurk) using the following inclusion criteria: 1) adults in the United States (US); 2) MTurk approval rate over 98\%; 3) at least 1000 previously approved tasks; 4) have not yet taken this task. We also followed these exclusion criteria: 1) not fluent in English; 2) any auditory and/or visual impairment due to the design of the study stimuli. In total, 525 participants accepted the task, and of those, 471 finished the full study. We only included participants who successfully finished the full study in the analysis. Because the estimated time needed to complete the study was 10 minutes, and we aimed to target a $15$ US dollar hourly rate, participants were compensated with US\$2.5. Demographic details about the participants are found in Table A of the supplemental material.

\subsubsection{Phase 2: In-Person Study for Personalization Evaluation}

As shown in Table B of the supplemental material, in Phase 2, we used university mailing lists to recruit 94 university students for the in-person study. The inclusion criterion was being enrolled as a university student, and we used the same exclusion criteria as in Phase 1. Overall, 94 participants completed the full study and were included in the data analysis. Participant ages in Phase 2 ranged from 17 to 32 years old (M = 22.6, SD = 2.7). Demographic details about the participants are found in Table B of the supplemental material. Because the estimated time needed for the Phase 2 study was 20 minutes, and we aimed to target a $15$ US dollar hourly rate, the participants were compensated with US\$5. 

In both Phases 1 and 2, the participants were recruited and randomly assigned to each study condition. The order of voice presentations was also randomized in Phase 2. Because of the randomized assignment, these conditions were not perfectly balanced with respect to age, gender, profession, and educational background.

\subsection{Experiment Setup}

\subsubsection{Audio Selection}
In both Phase 1 and 2, the 2-minute mindfulness practice instructions were recorded using the following 5 voices. We first recorded the instructions for feminine and masculine voices from human mindfulness therapists with a similar pace and tone. For synthesized voices, we included 3 Amazon AWS Polly TTS voices: Sally (feminine), Matthew (masculine) and Justin (child-like). We chose Sally and Matthew because they were the top-rated feminine and masculine voices for the long-form reading task in a large-scale evaluation study~\cite{cambre2020choice}. We also included a child-like voice (Justin) because that voice has been successfully used in past socially assistive robots deployed with children~\cite{clabaugh2019long}. Default settings of voice features were used to synthesize the 2-minute audio recordings for the 3 TTS voices. Both the TTS and human therapists' voices have approximately the same words per minute. 
\subsubsection{Physical Embodiment} As shown in Figure~\ref{fig:embodi}, three physical embodiment conditions were studied: 1) no agent: an audio player with an mp3 icon only; 2) conversational agent: Alexa Echo Dot, a physically embodied smart speaker not capable of any physical movement; 3) socially assistive robot: Blossom robot, a physically embodied socially assistive robot capable of physical body movement and interaction. The Alexa Echo Dot was selected because it is one of the most popular smart speakers and works well with the Amazon Polly synthesized voices used in the study. Blossom was chosen because is a simple, engaging, and affordable 3D-printed socially assistive robot~\cite{suguitan2019blossom}. 

\subsubsection{Mindfulness Exercise Setup} In Phase 1, for the No Agent condition, as shown in Figure~\ref{fig:embodi}, the window of an mp3 player bar was displayed on the online survey during the mindfulness practice. For the practice guided by Alexa Echo Dot and socially assistive robot conditions, a video recording was played together with the audio instructions on the online survey. In Phase 2, participants listened to the voice through an mp3 player bar for the No Agent condition. For the two agent-based conditions in Phase 2, participants interacted with the physical agents.

\subsubsection{Personalization Panel} In Phase 2 in-person study, we introduced a web-based user interface panel that allowed participants to adjust 5 major controllable features offered by Amazon Polly's API: gender, accent, pitch, speed, and break. By selecting from a range of values using sliders on the web-based UI panel, participants could generate real-time audio clips based on the chosen settings of voice features. The final settings of voice features were used to synthesize the user-personalized TTS voice candidate to compare with the non-personalized baseline voices. The participants were asked to use this process of personalization repeatedly until they obtained the desired voice.

\subsection{Measures}

\subsubsection{Quantitative Data}
In both Phase 1 and 2, participants were prompted with the same set of 7-point Likert-scale quantitative survey questions about their perceived quality of the voices. We developed this set of quantitative survey questions based on the previous large-scale TTS evaluation work~\cite{bennett2005large,cambre2020choice}. This survey aims to evaluate the voices based on the following criteria: a) overall voice quality for mindfulness meditation; b) clarity; c) emotional expressiveness; d) naturalness. In Phase 2, we also logged each user's choices of voice features (gender, accent, pitch, speed, and break time) after they completed the process of fine-tuning the voice to match their preferences. 

\subsubsection{Qualitative Data}

In both phases, we asked participants to explain the reasoning behind their ratings in 2 to 3 sentences. The goal was to enable further qualitative analysis into why certain voices outperform others in the mindfulness context.




\begin{figure*}[t!]
  \includegraphics[width=\textwidth]{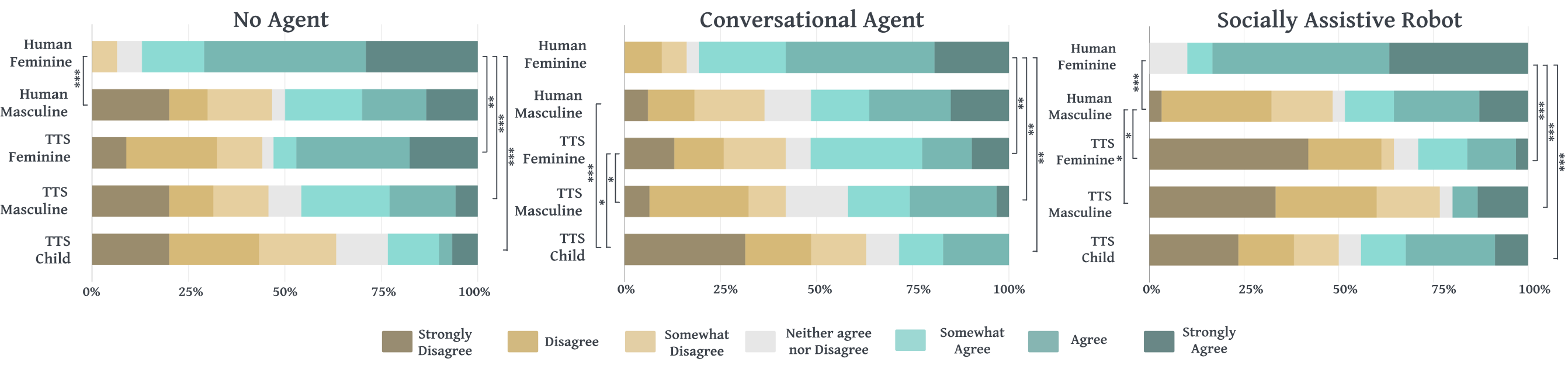}
  \caption{H1 (Phase 1): Voice evaluation in each embodiment condition}
  \label{fig:h1}
\end{figure*}
\begin{figure*}[t!]
  \includegraphics[width=\textwidth]{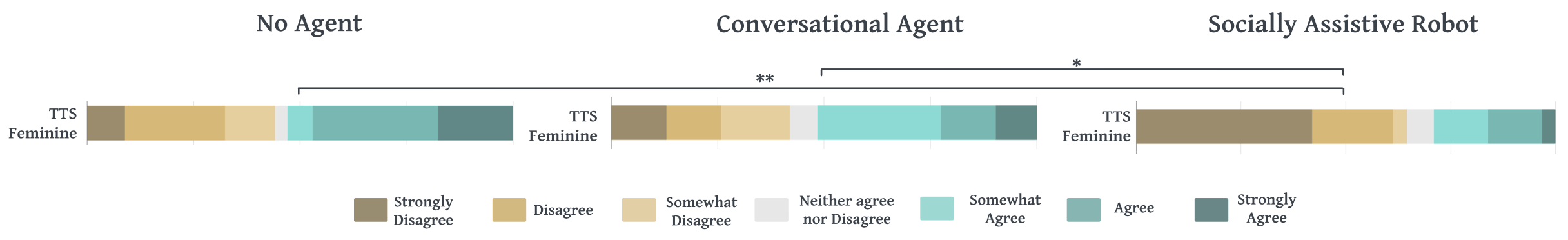}
  \caption{H2 (Phase 1): Voice evaluation in each embodiment condition}
  \label{fig:h2}
\end{figure*}

\begin{figure*}[t!]
  \includegraphics[width=\textwidth]{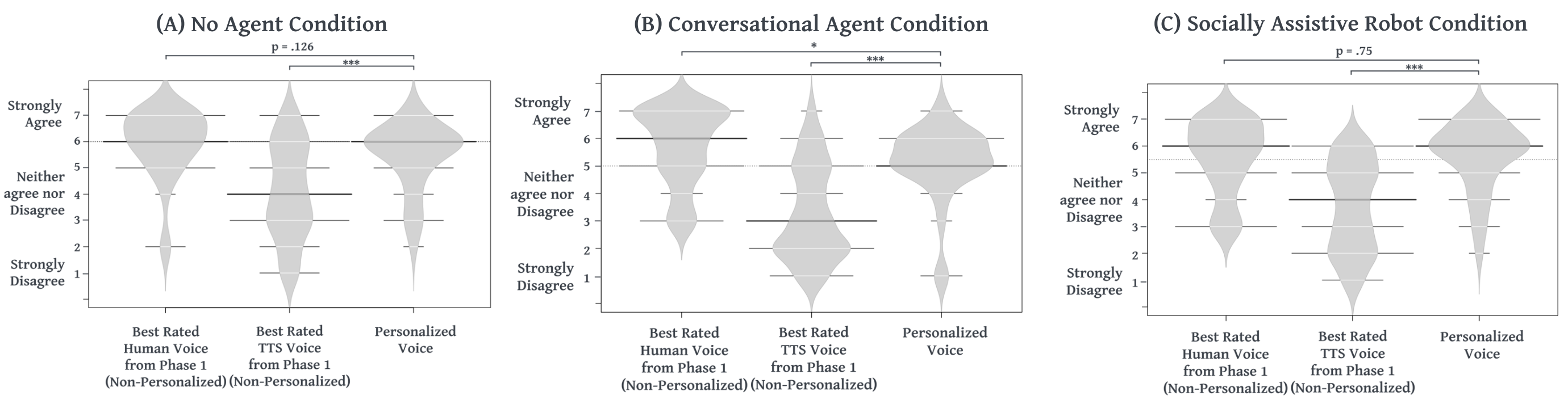}
  \caption{H3 (Phase 2): User-personalized vs. non-personalized voices in the three physical embodiment conditions: (A) no agent; (B) conversational agent; (C) socially assistive robot}
  \label{fig:h3}
\end{figure*}

\subsection{Analysis}
\label{sec:analysis}
This two-phase study measured the users' opinions of 3 TTS voices and 2 human voices(feminine/masculine/child-like) in 3 physical embodiment settings (no agent/conversational agent/socially assistive robot). The dependent variables of this study were: 1) the overall quality of the voice for mindfulness; 2) clarity of the voice; 3) emotional expressiveness of the voice; 4) naturalness of the voice. The dependent variables were rated on a 7-point Likert scale, and were coded from 1 to 7. The Shapiro–Wilk test was used to determine that the dependent variables were not normally distributed, so we used non-parametric tests for all of the analyses. We had two independent variables: voice type (five options) and embodiment type (three options). This resulted in 15 conditions to which participants were randomly assigned.

\paragraph{H1: voice evaluation between TTS and human therapists' voices} Because independent groups of participants were recruited for each voice condition~\cite{walpole1993probability}, the two-sided Mann-Whitney U test was used to test the difference between the quality of the TTS and human therapists' voices for each physical embodiment. To understand the differences among the 3 TTS voices, the two-sided one-way Kruskal-Wallis test was used. To compare between 2 TTS voices, the two-sided Mann-Whitney U test was also used. 

To further understand what contributed to the user-perceived overall quality of TTS voices for mindfulness meditation, ordinal logistic regression was used to establish the relationship between overall quality and specific voice characteristics, including: a) clarity, b) emotional expressiveness, and c) naturalness. The two-sided one-sample Wilcoxon signed rank test was also used to determine whether the median of users' ratings for these characteristics is statistically significantly equal to "Neither Agree nor Disagree (4)" on the 7-point Likert scale.

\paragraph{H2: voice evaluation between physical embodiment} Because separate groups of participants were also recruited independently for the three physical embodiments~\cite{walpole1993probability}, the two-sided one-way Kruskal-Wallis test was used to evaluate the differences among the 3 physical embodiment conditions for the same voice. To compare 2 physical embodiment conditions for the same voice, the two-sided Mann-Whitney U test was used. The factorial ANOVA was applied to test for the interaction effect between the voice and physical embodiment conditions.

\paragraph{H3: evaluation of user personalization} To test for differences between the user-personalized voice and the non-personalized voices~\cite{walpole1993probability}, because the same participants provided ratings for all three embodiments, the two-sided Wilcoxon signed-rank test was used. To better understand the distribution of the user-personalized voices, we used Principal Component Analysis (PCA) to apply dimension reduction to the five user-selected voice features (gender, accent, pitch, speed, and break time), and visualized the distribution, as shown in Figure A of the supplemental material.

For all the statistical tests, we applied Holm–Bonferroni correction to account for Type I errors in pairwise comparisons; $\alpha$ levels <.05*, <0.01**, and <.001*** were used to determine statistical significance.

\section{Results}

The analyses of H1 and H2 were conducted with the data from the Phase 1 MTurk study, and the analysis of H3 was conducted with the data from the Phase 2 in-person study.

\subsection{H1: Voice evaluation between TTS voices and human therapists' voices }

\subsubsection{No agent condition} As shown in Figure~\ref{fig:h1}, between the 2 human therapists' voices, the feminine human voice was rated significantly higher than the masculine human voice in user-perceived quality (U = 230.5, $p$ < .001). The better rated feminine human voice also was rated significantly higher than all the three TTS voices (feminine TTS: U = 334, $p$ = .009; masculine TTS: U = 212, $p$ < .001; Child-like TTS: U = 112.5, $p$ < .001). There was no significant difference found between the lower-rated masculine human voice and 3 TTS voices (feminine TTS: U = 451.5, $p$ = .848; masculine TTS: U = 493, $p$ = .67; Child-like TTS: U = 349, $p$ = .39). There was also no significant difference found between the 3 TTS voices (H(2) = 5.404, $p$ = .067).

\subsubsection{Conversational agent condition} As shown in Figure~\ref{fig:h1}, there was no significant difference found between the 2 human therapists' voices (U = 372.5, $p$ = .57). The feminine human voice was rated significantly higher than all the three TTS voices (feminine TTS: U = 282.5, $p$ = .008; masculine TTS: U = 260, $p$ = .03; Child-like TTS: U = 169, $p$ < .001). The masculine human therapist's voices was found no significant difference with two of the 3 TTS voices(feminine TTS: U = 449, $p$ = .395; masculine TTS: U = 425.5 , $p$ = .482), but significant higher than the child-like TTS voice (U = 287.5, $p$ = .003). 

Between the 3 TTS voices, a significant difference was found (H(2) = 7.309, $p$ = .026). Both the feminine (U = 328, $p$ = .038) and masculine (U = 462.5, $p$ = .042) TTS voice was rated higher than the child-like TTS voice, and no significant difference found between these two voices (U = 330.5, $p$ = .797).

\subsubsection{Socially assistive robot condition} As shown in Figure~\ref{fig:h1}, between the 2 human therapists' voices, the feminine human voice was rated significantly higher than the masculine human voice in user-perceived quality (U = 199.5, $p$ < .001). The better rated feminine human voice also was rated significantly higher than all the three TTS voices (feminine TTS: U = 91.5, $p$ < .001; masculine TTS: U = 123.5, $p$ < .001; Child-like TTS: U = 188.5, $p$ < .001). The lower-rated masculine human voice was also rated significantly higher than 2 of the 3 TTS voices (feminine TTS: U = 278, $p$ = .012 ; masculine TTS: U = 290, $p$ = .02), but not significantly different from the child-like TTS (U = 459.5, $p$ = .368). There was also no significant difference found between the 3 TTS voices (H(2) = .936 , $p$ = .14).

\subsubsection{Relationship between quality and voice characteristics for TTS} We conducted ordinal logistic regression between the overall quality of the TTS voices and the ratings of the 3 voice characteristics--clarity, emotional expressiveness, and naturalness--using the combined data from different voice and physical embodiment conditions. We found a significant relationship between all 3 voice characteristics: a) clarity (z = 2.205, $p$ = .026); b) emotional expressiveness (z = 6.796, $p$ < .001); and c) naturalness (z = 7.871, $p$ < .001). In addition, user-perceived ratings of voice clarity were significantly higher than neutral (neutral=4 on the Likert scale) (Mean = 5.35, Median = 6, $p$ < .001). However, the ratings for both emotional expressiveness (Mean = 3.69, Median = 4, $p$ = .004) and naturalness (Mean = 3.48, Median = 3, $p$ < .001) were significantly lower than neutral.

\subsection{H2: Voice evaluation between physical embodiments (no agent vs. conversational agent vs. socially assistive robot)}

We found a significant effect of physical embodiment in user-perceived quality for 1 of the 5 voices, which is the feminine TTS voice (H(2) = 10.593, $p$ = .005). More specifically, the feminine TTS voice in socially assistive robot condition was rated significantly lower than the No Agent (U = 296.5, $p$ = .002) and conversational agent (U = 312, $p$ = .016) conditions. There was no significant difference between the No Agent and conversational agent conditions (U = 550, $p$ = .889). We also found a significant interaction effect between the voice and embodiment ($F$ = 2.95, $p$ = .0032). 

There was no significant difference between 3 physical embodiment settings for both the feminine (H(2) = 4.972, $p$ = 0.083) and masculine (H(2) = 0.892, $p$ = .64) human therapists' voices. We also did not find significant difference between the 3 physical embodiments for 2 of the 3 TTS voices (masculine TTS:H(2) = 5.777, $p$ = .056; Child-like TTS: H(2) = 2.955,$p$ = .228).

\subsection{H3: Evaluation of user personalization (user-personalized vs. non-personalized voices)}

\subsubsection{No agent condition} As shown in Figure~\ref{fig:h3} (A), the user-personalized voice was rated significantly higher than the best rated TTS voice from Phase 1 (r = 0.465, $p$ < .001). We also found that there is no significant difference between the user-personalized voice and the best-rated human voice from Phase 1 (r = 0.194, $p$ = .126).

\subsubsection{Conversational agent condition} As shown in Figure~\ref{fig:h3} (B), the user-personalized voice was rated significantly higher than the best rated TTS voice from Phases 1 (r = 0.505, $p$ < .001). We found that the best-rated human voice from Phase 1 was rated significantly higher than the user-personalized voice (r = 0.269, $p$ = .034).

\subsubsection{Socially assistive robot condition} As shown in Figure~\ref{fig:h3} (C), the user-personalized voice was rated significantly higher than the best rated TTS voice from Phase 1 (r = 0.485, $p$ < .001). We also found that there is no significant difference between the user-personalized voice and the best-rated human voice from Phase 1 (r = 0.04, $p$ = .75).

As shown in Figure A of the supplemental material, we reduced the user-personalized choices of 5 voice features (gender, accent, pitch, speed, and break time) into 2 dimensions using Principle Component Analysis (PCA). This visualization shows that there are no obvious clusters of user preferences, and there is a large variance among users' preferred voices for mindfulness meditation.

\section{Discussion}

This work evaluated \textcolor{red}{}{user-perceived quality of} three state-of-the-art TTS voices on three embodiment conditions (H1, H2) applied to mindfulness practice, and explored the effect of TTS voice personalization (H3). Relative to each hypothesis, we found that: 1) all TTS voices were perceived significantly worse for mindfulness than the best-rated human therapist's voice, despite being rated similarly to the other lower-rated human voice (therefore, H1 was supported); 2) physical embodiments showed effects on one of the TTS voices for mindfulness, but not for the other TTS voices or human voices (therefore, H2 was partially supported); and 3) user-personalized voices not only outperformed the best rated TTS voice from Phase 1, but also performed almost as well as human voices (therefore, H3 was supported).

\subsection{H1: The best-rated human voice outperformed all TTS voices}

As shown in Figure~\ref{fig:h1}, we found that the human feminine therapist's voice was rated highest across all physical embodiments, and it was rated higher than all the TTS voices. While past studies have shown TTS voices are highly effective in the context of reading~\cite{wagner2019speech,cambre2020choice}, our findings for the mindfulness context indicate that TTS voices may not perform as well in others, as yet untested application domains, highlighting the need for evaluation in each application domain. In addition, our results show that there is still room for improvement of TTS voices to reach human-level performance. To understand why, we also analyzed user-perceived levels of clarity, emotional expressiveness, and naturalness of TTS voices. Our results show that users were in general satisfied with TTS voice clarity, but gave significantly more negative ratings for TTS voice emotional expressiveness and naturalness, providing direction for future TTS research and development.

Interestingly, at least one TTS voice performed as well as the human masculine therapist's voice in all three physical embodiment conditions. This suggests that there may exist a large variance in human voice quality for the mindfulness context, and TTS voices may at least perform similarly to the lowest-ranked human voices for mindfulness meditation. Participants also reported that the human masculine therapist's voice sounded "too stiff", and that they would prefer a "softer" and "more smoothing" voice. This shows that certain features of the human voice (e.g., gender, pitch, accent) may give some human voices an advantage, but are difficult for people to control, further supporting the advantage of using TTS voices if those TTS voices can easily adapt to each user's preferences.

Of the three TTS voices, none performed significantly better than the others for any of the three embodiments. There was no significant difference between the three voices in the No Agent and socially assistive robot conditions. In the conversational agent condition, we found that both feminine and masculine TTS voices were higher rated than child-like TTS voices, but there was no clear winner between the feminine and masculine TTS voices. Despite the potential effects of societal bias on gendered voices reported in past work~\cite{crowelly2009gendered}, our results show that a feminine voice would not necessarily be preferred over a masculine voice, and that the benefit of finding the best one-size-fits-all TTS voice for a given use (e.g., mindfulness) may be marginal.

\subsection{H2: Physical embodiment had an effect on one TTS voice, but not the other TTS and human voices}

We found that the feminine TTS voice had a significantly lower rating in the socially assistive robot condition than in the No Agent and conversational agent conditions. This is somewhat surprising, because we expected the computer-synthesized TTS voice to align better with the robot's physical embodiment than the other voice options, and therefore to receive higher ratings of the user-perceived voice quality. We found the opposite to be the case. In our qualitative data analysis, participants reported that the TTS voice sounded "too robotic" and "emotionless". Interestingly, in other TTS conditions, participants rarely mentioned the word "robotic". This suggests that the alignment between the robot embodiment and voice may help to remind and amplify the user's dislike of the artificial sound of a TTS voice. Furthermore, participants reported that the sound of the robot's servomotors during movement made it "hard to concentrate" during mindfulness meditation. This shows that servomotor sounds could have indirectly negatively influenced user ratings for TTS voices as well, and better engineering solutions need to be developed to minimize this negative impact of servomotor sounds.

We did not find any significant differences among the physical embodiments for the other 2 TTS voices. This could be because the Phase 1 study was conducted in MTurk, wherein the effect of physical embodiment is weakened compared to in-person human-robot interaction. Furthermore, the specific choices of agents may have a significant influence on the results, and a different set of agents may have elicited more significant effects of physical embodiment on user-perceived quality of TTS voices. 

\subsection{H3: User-personalized TTS voices performed better than non-personalized TTS voices, and almost as well as human voices} 

As shown in Figures~\ref{fig:h3}, we found that personalization could be a simple and very effective tool to improve the user-perceived quality of TTS voices. In all three physical embodiment conditions, by allowing the participants to fine-tune voice features (gender, accent, pitch, speed, and break time), the user-personalized voice condition consistently outperformed the best-rated TTS voices from Phase 1. More surprisingly, in both audio and socially assistive robot conditions, we did not find any statistical difference between the user-personalized TTS voice and the best-rated human feminine therapist's voice from Phase 1. As seen in Figure A of the supplemental material, this shows that users may have very different voice expectations and preferences for the mindfulness context, due to diverse user backgrounds and past experiences, so the one-voice-fits-all approach does not work as well as user-personalized voices. In addition, as suggested in ~\cite{spinuzzi2005methodology}, personalizing features for TTS voices may serve as a simple participatory design step, so that users can develop a sense of ownership of the TTS voice and have more realistic expectations of computer-synthesized TTS voices.


\subsection{Limitations and Future Work}
There are several limitations of this study. Due to a combination of reasons, including the COVID-19 pandemic and limited resources, the Phase 1 study (N=471) was conducted on Amazon Mechanical Turk (MTurk) instead of in person. MTurk results are always subject to possible limitations in generalization to in-person interactions. Our recruitment methods could have introduced biases relative to generalizing to various cultures, regions, or countries. Because our study was conducted in English, we recruited participants from the US only for Phase 1 on MTurk. For the in-person Phase 2 study, we recruited participants through university mailing lists. In addition, because of the randomized assignment, our conditions were not perfectly balanced with respect to age, gender, ethnicity, profession, and educational background.

The goal of this work was to evaluate user-perceived quality of voices, toward informing work that will evaluate clinical effectiveness of mindfulness interventions delivered by user-preferred voices. Since this study was conducted with three specific choices of agents, future work will need to explore how these findings generalize across different types of agents.

\section{Conclusion}

This work contributes results and insights from a novel two-phase study that compared and evaluated user-perceived quality of state-of-the-art TTS voices with human therapists' voices in the context of mediating mindfulness meditation in three embodiment conditions (no agent/conversational agent/socially assistive robot). Our findings show that: H1) all TTS voices were perceived as significantly worse for mindfulness than the best-rated human therapist's voice, despite being rated similarly to the other lower-rated human voice; H2) being physically embodied in robots unexpectedly worsened the user-perceived quality for one of the TTS voices, suggesting the alignment between the robot embodiment and voice may help to remind and amplify the user’s dislike of the artificial sound of a TTS voice. In addition, the sound of robots' servomotors may also negatively influence users' ratings of TTS voices; and H3) user-personalized voices outperformed the best-rated TTS voice from Phase 1, and, more surprisingly, they performed almost as well as human voices. This suggests that personalization may be a simple but very effective participatory design step toward improving user-perceived TTS voice quality.




\section{Acknowledgement}
This research was supported by the National Science Foundation National Robotics Initiative 2.0 Grant NSF IIS-1925083.

\bibliographystyle{ACM-Reference-Format}
\balance
\bibliography{base}

\appendix

\end{document}